\documentclass[a4full]{article}
\usepackage[margin=1.3in]{geometry}

\usepackage{bm,natbib,amsmath,amsthm,amsfonts}
\newtheorem{res}{Result}
\usepackage{algorithm}
\usepackage{algpseudocode}

\usepackage[colorlinks = true,linkcolor = blue, citecolor = blue]{hyperref}

\title{Stream Sampling with Immediate Decision}
\author{Bardia Panahbehagh, Rapha\"{e}l Jauslin and Yves Till\'e}
\date{\today}

\def\ab{\textbf{a}}
\def\pib{\boldsymbol{\pi}} 
\begin{document}

\maketitle

\begin{abstract}	
The manuscript introduces a method to select a random sample from a stream by deciding on each sampling unit immediately after observing it. The process could be applied to unequal as well as equal probability sampling. The implementation is straightforward. Algorithm selects a unit in the sample based on a single condition. It is particularly effective to make direct decisions on stream data, despite the data arriving in groups or the stream being linear.
\end{abstract}

\section{Introduction}\label{sec:intro}
There are a few number of sampling methods that can be applied on unequal probability stream data. Systematic sampling with unequal proba\-bilities~\citep{mad:49} is one of them which leads to a low entropy design and then is not proper to perform a control~\citep{bust:til:2020}. Two alternative methods are order pivotal method~\citep{dev:til:98} and Deville's systematic sampling~\citep{til:06}. In former, pairs of units are processed according to their order in the data list and in later the population will be partitioned in many strata based on the first order inclusion probabilities and one sample from each stratum will be selected.

Pivotal method and Deville's systematic sampling are two different imple\-mentations of the same sampling design~\citep{cha:12}. A disadvantage of the pivotal method is that the decision to take the first unit may be made after examining a very large number of units in the sampling frame. Disadvantages of Deville's systematic method include too many calculations for each step and problems with receiving data in groups in the middle of the process. In addition to these two algorithms, there are reservoir methods to sample from a stream~\citep{cha:82, cohen2009stream, til:19a}. In such methods, the reservoir is updated each time a unit enters. Then final decision on certain units will then be made very late. In stream sampling, it is a convenient property to be able to take a decision on each unit immediately after observing it.


The proposed method allows us to make an immediate decision on the units after observing them, independently of the units to come. Then the method is quite appropriate for stream data and can be applied to unequal as well as equal probability inclusion probabilities.

Section~\ref{sec:notation} introduces the notations and the principal concept of sampling theory. In Section~\ref{win1}, we present a stream sampling with windows of size one. Section~\ref{IDS} extends the method to immediate decision. Section \ref{sec:conclu} gives a summary of the method results.

\section{Basic concept}\label{sec:notation}

Consider a finite population $U = \{1,2,\dots,N\}$. Let $\mathcal{S} = \{s | s\subset U\}$ denote the power set of $U$. A sampling design is defined by a probability distribution $p(.)$ on $\mathcal{S}$ such that
$$
p(s) \geq 0 \text{ for all } s\in \mathcal{S} \text{ and }\sum_{ s\in \mathcal{S}}p(s) = 1.
$$
A random sample $S$ is a random vector that maps elements of $\mathcal{S}$ to an $N$ vector of 0 or 1 such that $\textrm{P}(S = s) = p(s)$. Let define $a_k(S)$, for $k = 1,\dots,N$:
$$
a_k =
\left\{\begin{array}{lll} 1 & \text{ if } k\in S\\ 0 & \text{ otherwise} . \end{array} \right.
$$
Then a sample can be denoted by means of a vector notation:
$
\ab^\top = (a_1,$ $a_2, $ $\dots, $ $a_N).
$ For each unit of the population, the inclusion probability $0\leq\pi_k\leq 1$ is defined as the probability that unit $k$ is selected into sample $S$:
\begin{equation*}\label{eq:pik}
\pi_k = \textrm{P}(k \in S) = \textrm{E}(a_k) =  \sum_{s\in \mathcal{S} | k \in s} p(s), \text{ for all } k\in U.
\end{equation*}

Let $\pib^\top=(\pi_1,\dots,\pi_N)$ be the vector of inclusion probabilities. Then, $\textrm{E}({\ab})=\pib.$ In many applications, inclusion probabilities are such that samples have a fixed size $n$.
The sample is generally selected with the aim of estimating some population parameters. Let $y_k$ denote a real number associated with unit $k\in U$, usually called the variable of interest. For example, the total
$$
Y=\sum_{k\in U} y_k
$$
can be estimated by using the classical Horvitz-Thompson estimator of the total defined by
\begin{equation*}
\widehat{Y}_{HT} = \sum_{k\in U} \frac{y_k a_k}{\pi_k}.
\end{equation*}

\section{Stream Sampling with Windows of Size One}\label{win1}

Consider a population $U$ with
$\bm{\pi}=(\pi_1,\pi_2,\dots,\pi_N)^\top.$ Let $v_1\in U$ be the index that
\begin{equation}\label{con1}
\sum_{k=1}^{v_1-1}\pi_k< 1 \mbox{ and } \sum_{k=1}^{v_1}\pi_k \ge 1,
\end{equation}
then we split $\pi_{v_1}=\pi_{v_{11}}+\pi_{v_{12}}$ such that
\begin{equation*}
\sum_{k=1}^{v_1-1}\pi_k+\pi_{v_{11}}=1.
\end{equation*}
Now, consider the first window $w_1$ of length 1, as
$$
\bm{\pi}=(
\underbrace{
\pi_1,\;\;\;\pi_2,\;\;\;\dots,\;\;\;\pi_{v_1-2},\;\;\;\pi_{v_1-1},\;\;\;\pi_{v_{11}}
}_{w_1:\;\;\text{The first window of size one}}+\;\pi_{v_{12}},\;\;\;\pi_{v_1+1},\;\;\;\dots,\;\;\;\pi_N)^\top.
$$
Based on \eqref{con1}, if $\sum_{k=1}^{v_1}\pi_k > 1$, we name unit $v_1$ as the cross-border unit of the first window, otherwise, $w_1$ does not have cross-border unit.

 
 As the sum of inclusion probabilities within the window is one, if a unit is selected, to compensate the needed inclusion probabilities, all the other inclusion probabilities will be updated to zero.

For the first unit, its inclusion probability will be updated as
\begin{eqnarray}\nonumber
\pi^1_1=\left\{\begin{array}{ll}
1& \text{ with probability } \pi_1 \\
0 & \text{ with probability } 1-\pi_1,
\end{array}\right.
\end{eqnarray}
and for the other units inside this window as the size of window is one, it is possible to update them as
\begin{equation*}
\pi^1_k=\left\{
\begin{array}{ll}
\displaystyle \pi^1_k(0)= \pi_k+\frac{\pi_k}{1-\pi_1}\pi_1=\frac{\pi_k}{1-\pi_1} & \mbox{ if } \pi^1_1=0\\[4mm]
\displaystyle\pi^1_k(1)=\pi_k-\frac{\pi_k}{1-\pi_1}(1-\pi_1)=0 & \mbox{ if } \pi^1_1=1,
\end{array}
\right.
\text{ for } k = 2,\dots,v_{11},
\end{equation*}
which leads to $E(\pi^1_k|\pi_1)=\pi_k$ and for the rest of the population there is no need to update the inclusion probabilities
$$
\pi^1_k=\pi_k, \mbox{ for } k=v_{12},v_1+1,v_1+2,\dots,N.
$$
It is possible to continue the process of updating inclusion probabilities inside $w_1$, until all the inclusion proba\-bilities inside are rounded to 0 and 1.

To continue, independent of the final decision on $w_1$, the next windows of size one should be considered as
\begin{multline*}
\bm{\pi}=(
\underbrace{
\pi_1,\pi_2,\dots,\pi_{v_1-2},\pi_{v_1-1},\pi_{v_{11}}
}_{w_1:\text{of size one }}\\+\underbrace{\pi_{v_{12}},\pi_{v_1+1},
\dots,\pi_{v_2-1},\pi_{v_{21}}}_{w_2:\text{of size one }}+\underbrace{\pi_{v_{22}},\pi_{v_2+1},\dots}_{w_3:\text{of size one\dots }},\dots)^\top,
\end{multline*}
based on the original inclusion probabilities. Then $v_\ell,$ as the cross-border of $w_\ell$, is a unit at the end of $w_\ell$ that helps to have a window of size one and a part of it will fall in $w_{\ell+1}.$ Formally with
$$F_0=0 \text{ and } F_j=\sum_{k=1}^{j}\pi_k,$$
unit $j$ is a cross-border unit if $F_j-F_{j-1}$ contains an integer number. For the last window, if it is not possible to make a window of size one, a phantom unit \citep{graf:mat:qua:til:12} will be added.

Now, as $v_{11}$ is not a real unit, depends on decision for this unit, the units inside $w_2$ will be initially updated as
\begin{equation}\label{edge}
\pi^*_{v_{12}}=\left\{
\begin{array}{ll}
\pi^{*1}_{v_{12}}=0 & \mbox{ if } \pi^v_{v_{11}}=1\\[2mm]
\displaystyle\pi^{*0}_{v_{12}}=\frac{\pi_{v_1}-\pi_{v_{11}}}{1-\pi_{v_{11}}} & \mbox{ if } \pi^v_{v_{11}}=0,
\end{array}
\right.
\end{equation}
it means, if $v_{11}$ is rounded to 1, then we consider unit $v_1$ as a selected unit and if $v_{11}$ is rounded to 0, then we do not make final decision about unit $v_1$ and postpone making decision about this unit later by updating $v_{12}$. Therefore we consider unit $v_1$ as a selected unit if at least one of $v_{11}$ or $v_{12}$ is selected. For the other units, they will be updated initially as
\begin{equation}\label{update}
\pi^*_k=\left\{
\begin{array}{ll}
\pi^{*1}_k=\frac{\pi_{k}}{1-\pi_{v_{12}}} & \mbox{ if } \pi^v_{v_{11}}=1\\[2mm]
\displaystyle\pi^{*0}_k=\frac{\pi_{k}-\pi^{*1}_{k}\pi_{v_{11}}}{1-\pi_{v_{11}}} & \mbox{ if } \pi^v_{v_{11}}=0,
\end{array}
\right.
\text{ for } k = v_1+1,\dots,v_{21}.
\end{equation}

The other units will not change. With this initially updating, the process respects the original inclusion probabilities and the size of $w_2$ will be preserved at one. 
After final decision about $v_{21}$, again initial updates should be imposed on the units inside $w_3$, and then the process continues until making the decision about the last unit in the population.

\section{Immediate Decision Sampling (IDS)}\label{IDS}


 Generally with stream sampling algorithm, to select the first unit, it is necessary to wait a certain number of units to compensate the information of the first unit. But here we show that based on the windows of size one, it is possible to make decision about the units immediately after observing them. Indeed relying windows of size one, there is no need to know the other units to compensate the information of the last decision.

For this purpose, let consider $w_\ell$ as the $\ell$th window containing $\pi_j$ or ending by a part of $\pi_j$, $\pi^{*}_j$ as the last update of $\pi_j$ before rounding it to 0 or 1, and finally $n_j$ as the size of selected sample before deciding about $\pi_j$.
\begin{res}\label{first}
The size-one window method is equivalent to the following principles. After observing unit $j$, if
\begin{itemize}
\item[I)] $j$ is not a cross-border unit,
\begin{enumerate}
\item if $n_j<\ell$,
$$
\pi^{*}_j=\frac{\pi_j}{1-(F_{j-1}-\lfloor F_{j-1}\rfloor)},
$$
\item if $n_j=\ell$,
$$
\pi^{*}_j=0.
$$
\end{enumerate}

\item[II)] $j$ is a cross-border unit,
\begin{enumerate}
\item if $n_j<\ell$,
$$
\pi^{*}_j=1,
$$
\item if $n_j=\ell$,
$$
\pi_j^{*}=\frac{\pi_j-\left\{1-(F_{j-1}-\lfloor F_{j-1}\rfloor)\right\}}{F_{j-1}-\lfloor F_{j-1}\rfloor}.
$$
\end{enumerate}
\end{itemize}
\end{res}
For proof see the Appendix.

Then based on IDS, for deciding about the observed unit, there is no need to know about unobserved units coming in future.
The four scenarios of IDS can be summarized in one condition as presented in Algorithm \ref{algo:1}. The following results, show that IDS (or windows of size one method), respects the first order and sum of inclusion probabilities and also IDS is an implementation that leads to the same design of Deville's systematic sampling.
\begin{res}\label{second}
With IDS, for all $j,\ell=1,2,\dots$,
\begin{itemize}
	\item[I)]  $\Pr(j\in S)=\pi_j,$
	\item[II)] $0\le \pi^{*0}_j,\pi^{*1}_j\le 1,$
	
	\item[III)] $\sum_{k=v_{\ell2}}^{v_{(\ell+1)1}}\pi^{*0}_k=\sum_{k=v_{\ell2}}^{v_{(\ell+1)1}}\pi^{*1}_k=1$.
\end{itemize}
\end{res}
For proof see the Appendix.
\begin{res}\label{deville}
IDS and Deville's systematic sampling (Algorithm \ref{algo:2}) are two implementations of the same design.
\end{res}
For proof see the Appendix.

Now based on Result \ref{deville} and \citep{dev:98a}, it is easy to calculate the second order inclusion probabilities.
\begin{res}\citep{dev:98a}\label{seconinclu}
Let $k$ and $k'$ be two distinct units in $U$.
\begin{itemize}
\item If $j$ and $j'$ are two non-cross-border units that belong to the same window, then
$$
\pi_{jj'}=0.
$$
\item If $j$ and $j'$ are two non-cross-border units that belong to distinct windows $\ell$ and $\ell'$, respectively, where $\ell<\ell'$ then
$$
\pi_{jj'}=\pi_j\pi_{j'}\{1-c(\ell,\ell')\}
$$
\item If $j=v_\ell$ and $j'$ is a non-cross-border unit that belong to distinct window $\ell'$, respectively, where $\ell<\ell'$ then
$$
\pi_{jj'}=\pi_{v_{\ell}}\pi_{j'}\{1-\frac{\pi_{v_{\ell2}}(1-\pi_{v_{\ell}})}{\pi_{v_\ell}(1-\pi_{v_{\ell2}})}c(\ell,\ell')\}.
$$
\item If $j'=v_{\ell'}$ and $j$ is a non-cross-border unit that belong to the window $\ell$, where $\ell\le\ell'$ then
$$
\pi_{jj'}=\pi_{j}\pi_{v_{\ell'}}\{1-\frac{(1-\pi_{v_{\ell'2}})(1-\pi_{v_{\ell'}})}{\pi_{v_{\ell'}}\pi_{v_{\ell'2}}}c(\ell,\ell')\}.
$$
\item If $j=v_\ell$ and $l=v_{\ell'}$, where $\ell<\ell'$ then
$$
\pi_{jj'}=\pi_{v_{\ell}}\pi_{v_{\ell'}}\{1-\frac{\pi_{v_{\ell2}}(1-\pi_{v_{\ell'2}})(1-\pi_{v_{\ell}})(1-\pi_{v_{\ell'}})}{\pi_{v_\ell}\pi_{v_{\ell'}}\pi_{v_{\ell'2}}(1-\pi_{v_{\ell2}}))}c(\ell,\ell')\},
$$
\end{itemize}
where
$$
c(\ell,\ell')=\Pi_{r=\ell}^{\ell'-1}c_r, c_r=\frac{\pi_{v_{r1}}\pi_{v_{r2}}}{(1-\pi_{v_{r1}})(1-\pi_{v_{r2}})},
$$
and with $c(\ell,\ell)=1.$
\end{res}

Then with a fixed order of the population units, IDS can be considered as a simpler version of Deville's method. In IDS, according to Algorithm \ref{algo:1}, there is no need investigate if a unit is a cross-border unit or not and also there is no need to generate random variables from different distribution for each windows.

\begin{algorithm}[htb!]
\caption{Algorithm of Immediate Decision}\label{algo:1}
\begin{algorithmic}
\State Initialize with $F=0$, $n=0$ and $s=\{\}$,
\State After receiving unit $j$ in a stream follow the following steps,
\begin{enumerate}
\item Generate $u$, a realization of a uniform random variable in $[0,1]$,
\item Set $F_1=F$, $F=F+\pi_j$, $\alpha=\lceil F\rceil-\lfloor F_1 \rfloor-1$, $\beta=\lceil F\rceil-n$ and $m=F_1-\lfloor F_1\rfloor$
\item If
$$
u\le \min(\beta,1)\frac{\pi_j-\alpha(2-\beta)(1-m)}{(1-\alpha)(1-m)+\alpha\left\{(2-\beta)m+\pi_j(\beta-1)\right\}}
$$
then
$$
s=s\cup j, n=n+1.
$$
\end{enumerate}
\end{algorithmic}
\end{algorithm}

\begin{algorithm}[htb!]
\caption{Deville's systematic procedure}\label{algo:2}
\begin{algorithmic}
\State Generate $u_1$, a realization of a uniform random variable in $[0,1]$. Unit $j$ such that $F_{j-1} \le u_1< F_j$ is selected.
\For {$i=2,\dots,n$},
\State Let $\ell$ be the cross-border unit such that $F_{\ell-1}\le i-1 <F_{\ell}$
\If {unit $\ell$ is selected at step $i-1$}
\State \begin{eqnarray}\nonumber
f(x)=\left\{\begin{array}{ll}
\frac{1}{\lceil F_\ell \rceil-F_\ell}& \text{ if } x\ge F_\ell-\lfloor F_\ell\rfloor \\
0 &  \text{ if } x<F_\ell-\lfloor F_\ell\rfloor,
\end{array}\right. x\in [0,1[.
\end{eqnarray}
\Else
\State \begin{eqnarray}\nonumber
f(x)=\left\{\begin{array}{ll}
1-\frac{(\lceil F_{\ell-1}\rceil-F_{\ell-1})(F_\ell-\lfloor F_\ell \rfloor)}{\left\{1-(\lceil F_{\ell-1}\rceil-F_{\ell-1})\right\}\left\{1-(F_\ell-\lfloor F_\ell \rfloor)\right\}}& \text{ if } x\ge F_\ell-\lfloor F_\ell\rfloor \\\\
\frac{1}{1-(\lceil F_{\ell-1}\rceil-F_{\ell-1})} &  \text{ if } x<F_\ell-\lfloor F_\ell\rfloor,
\end{array}\right. x\in [0,1[
\end{eqnarray}
\EndIf
\State Generate $u_i$, a random variable with density $f(x)$.
\State Unit $j$ such that $F_{j-1}\le u_i+i-1<F_j$ is selected
\EndFor
\end{algorithmic}
\end{algorithm}



\section{Summary}\label{sec:conclu}
In this paper, we proposed a sampling method that allows the selection of sample units immediately after observing them without any information about the upcoming data. This method satisfies exactly equal and unequal inclusion probabilities. The proposed algorithm gives a straightforward imple\-mentation of Deville's (ordered pivotal) sampling design.

%
%

\section*{Appendix}

\subsection*{Proof of Result \ref{first}}

Proof is based on Equations \eqref{edge} and \eqref{update}. Consider the unit $j$, inside $w_2$
\begin{itemize}
\item[I)] $j$ is not an cross-border unit,
\begin{enumerate}
\item if $n_j<\ell$ and $\pi^v_{v_{11}}=1$, then according to $\eqref{update}$ we have
$$
\pi^{*}_j=\frac{\frac{\pi_j}{1-\pi_{v_{12}}}}{1-\sum_{i=v_1+1}^{j-1}\frac{\pi_{j}}{1-\pi_{v_{12}}}}=\frac{\pi_j}{1-(F_{j-1}-\lfloor F_{j-1}\rfloor)},
$$
\item if $n_j=\ell$, according to windows of size one, this unit will not be selected.
\end{enumerate}
\item[II)] $j$ is a cross-border unit,
\begin{enumerate}
\item if $n_j<\ell$, according to windows of size one, this unit will be selected with probability one.
\item if $n_j=\ell$,
$$
\pi^{*}_j=\frac{\frac{\pi_{j}-\pi^{*1}_{j}\pi_{v_{11}}}{1-\pi_{v_{11}}}}{1-\frac{\pi_{v_1}-\pi_{v_{11}}}{1-\pi_{v_{11}}}-\sum_{i=v_1+1}^{j-1}\frac{\pi_{j}-\pi^{*1}_{j}\pi_{v_{11}}}{1-\pi_{v_{11}}}},
$$
and then replacing $\pi^{*1}_j$ by $\pi_j/(1-\pi_{v_{12}})$ we have
$$
\pi^{*}_j=\frac{\pi_j}{1-(F_{j-1}-\lfloor F_{j-1}\rfloor)}.
$$
\end{enumerate}
For the other windows the proof is the same.
\end{itemize}

\subsection*{Proof of Result \ref{second}}

\subsubsection*{I):}

Inside $w_1$, for the non-cross-border units, we have
\begin{eqnarray}
\Pr(j\in S)&=&\Pr(1\notin S, 2\notin S,\dots,j-1\notin S, j\in S)\nonumber\\ &=&\left(1-\pi_1\right)\left(1-\frac{\pi_2}{1-\pi_1}\right)\left(1-\frac{\pi_3}{1-\pi_1-\pi_2}\right)\dots\nonumber\\
&\times&\left(1-\frac{\pi_{j-1}}{1-\sum_{k=1}^{j-2}\pi_k}\right)\left(\frac{\pi_j}{1-\sum_{k=1}^{j-1}\pi_k}\right)\nonumber\\
&=&\pi_j.\nonumber
\end{eqnarray}
For $j=v_1$,
\begin{eqnarray}
\Pr\left(v_1\in S\right)&=&\Pr\left(\text{$v_1$ is selected in $w_1$}\right)\nonumber\\
&+&\Pr\left( \text{$v_1$ is not selected in $w_1$} \mbox{ and } \text{$v_1$ is selected in $w_2$}\right)\nonumber\\
&=&\pi_{v_{11}}+\left(1-\pi_{v_{11}}\right)\left(\frac{\pi_{v_1}-\pi_{v_{11}}}{1-\pi_{v_{11}}}\right)=\pi_{v_1}.\nonumber
\end{eqnarray}
For unit $j$ inside $w_2$ ($i=j+v_1$),
\begin{eqnarray}
\Pr\left(i\in S\right)&=&\Pr\left(\text{$v_1$ is selected in $w_1$} \mbox{ and } \text{$i$ is selected in $w_2$}\right)\nonumber\\
&+&\Pr\left( \text{$v_1$ is not selected in $w_1$} \mbox{ and } \text{$i$ is selected in $w_2$}\right)\nonumber\\
&=&\pi_{v_{11}}\left(1-\pi^{*1}_{v_1+1}\right)\left(1-\frac{\pi^{*1}_{v_1+2}}{1-\pi^{*1}_{v_1+1}}\right)\dots\left(\frac{\pi^{*1}_{v_1+j}}{1-\sum_{k=v_1+1}^{v_1+j-1}\pi^{*1}_k}\right)\nonumber\\
&+&\left(1-\pi_{v_{11}}\right)\left(1-\pi^{*0}_{v}\right)\left(1-\frac{\pi^{*0}_{v_1+1}}{1-\pi^{*0}_{v_1}}\right)\dots\left(\frac{\pi^{*0}_{i}}{1-\sum_{k=v_1}^{v_1+j-1}\pi^{*0}_k}\right)\nonumber\\
&=&\pi_{v_{11}}\frac{\pi_i}{1-\pi_{v_{12}}}+\left(1-\pi_{v_{11}}\right)\frac{\pi_{i}-\frac{\pi_i}{1-\pi_{v_{12}}}\pi_{v_{11}}}{1-\pi_{v_{11}}}=\pi_i.\nonumber
\end{eqnarray}

\subsubsection*{II):}
Inside $w_1$, for the non-cross-border units, as the size of $w_1$ is 1,
$$1-\sum_{k=1}^{j-1}\pi_k=\sum_{k=j}^{v_{11}}\pi_k,
$$
and then $0\le \pi^*_j\le 1$. For all the non-cross-border units of $w_2$, based on the same reason, again as the size of $w_2$ is 1, $0\le \pi^{*1}_j\le 1$. For the first cross-border unit, as $\pi_{v_1}\le 1$ then $0\le \pi^{*1}_{v_{12}} \le 1$. Also $0\le \pi^{*0}_{j}\le 1$ in $w_2$, if and only if
\begin{eqnarray}
\pi_{j}(1-\pi_{v_1})\ge 0, \mbox{ and } \pi_j\le 1+\frac{\pi_{v_{11}}\pi_{v_{12}}}{1-\pi_{v_1}},\nonumber
\end{eqnarray}
that both of them are obviously satisfied.

\subsubsection*{III):}
For $w_1$ it is obvious. For $w_2$, with $\pi^v_{v_{11}}=1$,
$$
\sum_{k=v_{12}}^{v_{21}}\pi^{*1}_k=0+\frac{\sum_{k=v_1+1}^{v_{21}}\pi_{k}}{1-\pi_{v_{12}}}=\frac{1-\pi_{v_{12}}}{1-\pi_{v_{12}}}=1,
$$
and with $\pi^v_{v_{11}}=0$,
\begin{eqnarray}
\sum_{k=v_{12}}^{v_{21}}\pi^{*0}_k&=&\frac{\pi_{v_1}-\pi_{v_{11}}}{1-\pi_{v_{11}}}+\frac{\sum_{k=v_1+1}^{v_{21}}\pi_{k}-\pi_{v_{11}}\sum_{k=v_1+1}^{v_{21}}\pi^{*1}_{k}}{1-\pi_{v_{11}}}\nonumber\\
&=&\frac{\pi_{v_{12}}}{1-\pi_{v_{11}}}+\frac{(1-\pi_{v_{12}})-\pi_{v_{11}}\times 1}{1-\pi_{v_{11}}}=1.\nonumber
\end{eqnarray}
For the other windows, the proof is the same as $w_2$.
\subsection*{Proof of Result \ref{deville}}

The structure of the population and cross-units inside both methods are the same, the updating principle of the Deville's method is the same as Equations \eqref{edge} and \eqref{update}. In Deville's method, consider the two first windows,
\begin{itemize}
\item[I)] $j$ is not an cross-border unit,
\begin{enumerate}
\item If the cross-unit is selected inside the previous window, then
\begin{eqnarray}
P(j\in S)&=&\int\limits_{F_{j-1}}^{ F_{j}}f(x)dx=\int\limits_{F_{j-1}}^{ F_{j}}\frac{1}{\lceil F_{v_1}\rceil-F_{v_1}}dx=\frac{1}{\lceil F_{v_1}\rceil-F_{v_1}}\pi_j\nonumber\\
&=& \frac{1}{1-\pi_{v_{12}}}\pi_j\nonumber
\end{eqnarray}
which is equivalent to the first term of \eqref{update},
\item If the cross-unit is not selected inside the previous window, then
\begin{eqnarray}
P(j\in S)&=& \int\limits_{F_{j-1}}^{ F_{j}}1-\frac{(\lceil F_{v_1-1}\rceil-F_{v_1-1})(F_{v_1}-\lfloor F_v \rfloor)}{\left\{1-(\lceil F_{v_1-1}\rceil-F_{v_1-1})\right\}\left\{1-(F_{v_1}-\lfloor F_v \rfloor)\right\}}dx\nonumber\\
&=&\frac{1-\pi_v}{(1-\pi_{v_{11}})(1-\pi_{v_{12}})}\pi_j\nonumber
\end{eqnarray}
which is equivalent to the first term of \eqref{update},
\end{enumerate}
\item[II)] $j$ is an cross-border unit ($j=v_1$),
\begin{enumerate}
\item If the cross-unit is selected inside the previous window, then the method ignore the second part of $v_1$, i.e. ($\pi^*_{v_{11}}$=0), which is equivalent to the first term of \eqref{edge},
\item If the cross-unit is not selected inside the previous window, then
\begin{eqnarray}
P(j\in S)&=&\int\limits_{\lfloor F_{j}\rfloor}^{ F_{j}}\frac{1}{1-(\lceil F_{v_1-1}\rceil-F_{v_1-1})}dx\nonumber\\&=&\frac{1}{1-(\lceil F_{v_1-1}\rceil-F_{v_1-1})}(F_{j}-\lfloor F_{j}\rfloor)\nonumber\\
&=& \frac{1}{1-\pi_{v_{11}}}\pi_{v_{12}}=\frac{\pi_{v_1}-\pi_{v_{11}}}{1-\pi_{v_{11}}}\nonumber
\end{eqnarray}
which is equivalent to the second term of \eqref{edge},
\end{enumerate}
\end{itemize}
For the other windows the proof is the same.

Now if $s$ is a fixed sample and $p_I(.)$ and $p_D(.)$ are the designs of IDS and Deville's method respectively, as all the units inside $s$ have to be selected under the same principal in the both method, then
$p_I(s)=p_D(s).$

\end{document}